# Physical Consequences of a Momenta-Transfering Particle Theory of Induced Gravity and New Measurements Indicating Variation from Inverse Square Law at Length Scale of .1 mm: Statistical Time Properties of Gravitational Interaction and Analysis Thereof

Gary Christopher Vezzoli, Ph.D., Institute for the Basic Sciences
51 Park Drive, Suite 22, Boston, MA 02215;
Arlington High School Physics Department,
Arlington, MA 02476; April 2nd, 2001

Abstract

This work presents physical consequences of our theory of induced gravity (Ref.1) regarding: 1) the requirement to consider shape and materials properties when calculating graviton cross section collision area; 2) use of Special Relativity; 3) implications regarding the shape of cosmos; 4) comparison to explanations using General Relativity; 5) properties of black holes; 6) relationship to the strong force and the theorized Higgs boson; 7) the possible origin of magnetic attraction; 8) new measurements showing variation from gravitational inverse square behavior at length scales of 0.1 mm and relationship to the Cosmological constant, and proof of the statistical time properties of the gravitational interaction.

*1) Cross-Section Scattering Properties*:

In our theoretical work on induced gravity, employing omni-directional impacting neutrinos as the particle-momenta source and employing a shadowing physical model (1) we showed quantitatively that G is a *function of collision phenomena* rather than the constant postulated by Newton. This function has direct dependence on the neutrino collision/capture cross-section., and therefore must have dependence on *shape* and *materials properties* of the test object.  Each of the three neutrinos – tau, mu, and electron – is associated with a fundamental quark and lepton.  The electron neutrino is associated with the u (up) and d (down) quark, and therefore is interactive with the proton and when captured (to be then removed from gravitational momenta transfer in an otherwise incident momenta transfer on a subsequent nucleon) the electron neutrino induces conversion to a neutron.  In the analysis of induced gravity given in Ref. 1, we show that the expression for the scattering cross-section is extremely complicated, and introduces a correction factor to the inverse square law of Newton.  The evaluating of the scattering parameter (sigma) requires significant knowledge of the interactions of neutrinos with fundamental particles, this being a further motivation for work being conducted at CERN and at the Fermi Laboratory.

The neutrino detecting studies reported from Super K in Japan have indicated a substantial difference between neutrinos observed directly from the atmosphere and those observed from the path through the earth (the shadowed side), and a 2:1 ratio of tau neutrinos to electron neutrinos.  Interpretation of statistical studies have been employed to indicate that incoming neutrinos may be coupled and become oscillatory between the tau and the electron flavors, and thus the neutrinos are interpretive to have mass in order to satisfy this coupling requirement.

Shape will affect gravitational properties because of area-to-volume considerations.  Furthermore, some materials have a vast preference of absorbing neutrinos as compared to other materials, such as chlorine which undergoes transmutation to argon upon neutrino capture and is utilized as a neutrino detector in experiments aimed at studying the neutrino's all-important property of bona fide mass.  Based on the above, in a free fall experiment, it would be predicted by our model that objects of the identical mass value, but of different shape and chemical composition (differing in their nuclear chemistry regarding neutrino capture), will fall *equal* distances in theoretically and ideally *unequal* times, contrary to Galileo's interpretation, yet in consonance with relationship to some modern free fall experiments carried out early in the last decade.

If the above inferences are validated, then the gravity particle behavior must depend upon the weak force that governs radioactive decay and thus on the electro-weak union (or the particle that is parent to the weak interaction and the electromagnetic interaction).

2. *Use and Source of Special Relativity*

In the theory of induced gravity, it is essential to assume that the incident momenta transferring particles travel at the velocity of light, otherwise (without invoking Special Relativity), the drag force that would be caused by these particles on a planet, such as Earth, would create a very quick instability in the planet's orbit and engender engulfing by the Sun in short order (days). In a momentum-transferring gravitational particle, the origin of such Relativistic notions as the contraction of length and the dilation of time, then take on a *physical* explanation provided that it is the velocity of that particle which is the key and limiting factor, not the velocity of light. (In the subject physical model for induced gravity there appears no a priori reason for the velocity of *light* to be a physically limiting parameter). A linear object, such as a rod, that is traveling at c = velocity of the impacting neutrino will then suffer a stream of direct collisions in which the rod is moving toward a flux of gravity-carrying neutrinos and away from a reverse-direction flux of such particles. (This thought-experiment served as the basis of our earlier conceptualizing of the neutrino as the source of the property of inertia give in Ref. 1). The line of centers between atoms of the rod and individual neutrinos from the opposite direction in which the rod is moving would normally be thought to decrease at a rate of 2c, and the collision impact would have to *reduce* the length of the rod via compression and ultimately via virtual coalescing of electron clouds.  The shortened rod will then have a lower cross-section for neutrino capture in the direction of its length (from the physical and the nuclear chemistry standpoints) and the value of G will change which will cause a decrease in the local value of the acceleration due to gravity.  Any calculation such as that for the timing period of a pendulum that involves g in the denominator will diverge, and time will thus appear to dilate.  For a cogent physical picture, the mathematical transformations employed by Einstein in the theory of Special Relativity seemingly would have to substitute the velocity of the neutrino (which is the same magnitude as the velocity of light)  as the key reference constant, and abandon the mass transformation.

The neutrino flux model for the source of gravity could then argue that one must again consider a medium akin to an ether, the neutrinos that pervade the universe being that ether.  If the neutrino is moving, as postulated herein,  at the speed of c, then an ether composed of neutrinos would introduce no drag force on the planets, therefore not interfering with the gravitational explanation of their motion.  Einstein's theory of Special Relativity arose from his considerations of the theory of electromagnetic radiation and the uncommon property that electromagnetic waves propagate in a vacuum.  Now it is understood that there is no such perceived vacuum but instead, everywhere, a high density flux of neutrinos that could be postulated in order to explain gravity on a momenta basis rather than a field basis.  An omni-directional neutrino flux, postulated as a quasi ether, with no net drift velocity in any direction would yield the same results relative to the Michelson-Morley experiment as the interpretation of the experiment, namely, that there was no indication of an ether (assumed in the Michelson Morley analysis to be moving at some velocity, v).  The reason for this is that a spinning and orbiting earth would neither move into or away from a net flux of *omni-directional neutrinos*. If the postulate that light travels at a fixed speed through the ether is correct, then a critical inference in the reasoning at the heart of the interpretation of the Michelson-Morley experiment – namely, the expectation that if an observer is moving in the direction of the light ("trying to catch up with the light" that is emanated from a point source), the light would *appear* to that observer to be traveling more *slowly* than the appearance to an observer moving away from the direction of the light – would be as tenable for the above proposed "ether" as it would be for no ether whatsoever. However, according to the impinging particle induced gravity model and theory, the earth is orbiting *into* an imbalanced neutrino flux relative to where it is moving *from*, even though it is orbiting at only 0.001 times the speed of light. Therefore, light moving into the tangent of the direction of the earth's orbit is encountering a greater density of gravity-impinging particles than light moving in the opposite direction or at right angles. Ideally, then, there would have to be a fringe shift in the perfect Michelson Morley experiment. However, the magnitude of the ratio of the small neutrino-nucleon scattering cross-section to the *much* smaller neutrino-photon scattering cross-section (a ratio of about $1 \times 10^{30}$ ) would severely limit the practical feasibility of any fringe shift measurement.

Based on the above, the significance of a neutrino "ether" may tentatively seem to be in terms of the medium that actually may be required for the propagation of light! This seems puzzling because the photon-neutrino scattering interaction is believed to have such a low cross-section, however, significant gravity-bearing-particle/photon interaction may not be necessary in an environment of a very high density of neutrino flux. . [If the neutrino flux were not strictly omni-directional in some regions of the universe, such as near the expanding edges, then if neutrinos were a necessary medium for the propagation of light, a Michelson-Morley in that part of the universe may be able to detect a fringe shift].

3. Shape of Cosmos Inferences

Since a cosmological topology that can be believed to adhere to the laws of physics ("laws" seemingly inherently statistical in nature) between times of birth of a universe (big bang) or death of a universe (perhaps big "thud" or collapsing back on itself), then it seems that a "surviving" universe would contain constituent galaxies that would evolve to be cosmologically so shaped that overall gravitational particle absorption properties would act (as a lowest free energy tendency) to stave off gravitational collapse as much as possible, and thus not be spherical but laminar.

In a gravity-bearing external particle model, a final termination of a universe due to collapse could be caused by a net external particle flux in one direction which overbalanced the momentum and/or kinetic energy of an outward expanding universe in the opposite direction. Cosmological topology would then refer to the form revealed by the mapping of all mass other than neutrinos. This would be expected to show an irregularly shaped form which convoluted back on itself in a high entropy disordered manner leaving huge cavities (akin perhaps to a piece of slag residue from a steel-producing furnace), this in accord with modern topological views of cosmos but for differing reasons. A spherically symmetric cosmos could simply not resist gravitational forces because of relatively high neutrino-nucleon collision cross sections in all directions and would come crashing together in far less than 15 billion years, thus ultimately gravity serving as the source of extinction of one universe and perhaps re-birth of another via consequent big bang upon "final" convergence. On the other hand, on a small scale such as the shape of a droplet of water, the omni-directional nature of gravity-particle momenta transfer would render it approximately spherical

(being in consonance with a sphere having the smallest surface area for a given volume and best being able to "crouch" or protect itself against deformation due to gravity). An intermediate sized hollow body such as the red pepper would then reflect a long history of evolvement due to gravity, and does so in its elongation and near-flat-sided topology.

The above seems to imply that a surviving universe – one that can resist implosion to due gravitational momenta transfer – must have vast "pocket"-volumes of "empty space" and vast curved laminar (yet spatially continuous with other zones) regions of the probable existence of mass (such as spiral galaxies which are clearly *laminar,* but not necessarily nebulae or interstellar particles whose mass of each particle may be too small to tend to laminar geometry). This type of accretion shape argument relative to universal gravitation could then shed understanding on why there is a plane of the ecliptic in our *solar system*, yet no such plane in the quantum mechanical atom, since the electron-neutrino scattering cross-section must be extremely small so that the gravitational effect on the electron as part of the atom's electronic cloud would be virtually unobservable.

A universe that either in a random or forced manner had the property of large laminar-type topologies of the probable locus of spatial positions where mass existed would have a greater chance for survival against gravity than a universe of high spherical symmetry. The torus-shaped universe – the torus being a popular shape studied by topologists and of similar topological properties to the coffee-cup -- is commonly an example of a laminar shape with "length" and width greater than depth, and would be an example of a favorable shape to stave off gravitational collapse for the same reasons as those that favor spiral or cartwheel galaxy geometry..

A principle of nature seems to be emerging from the above line of deduction, namely that *systems in nature tend to evolve to configurations that best avoid detrimental gravitational interactions*. Free fall, then, occurs because it is less opposed gravitationally by collisions with gravity-bearing particles from the "unshadowed" direction (the direction from the atmosphere and space) and pushed or "boosted" by those particles coming "unshadowed" from the lithosphere. Laminar geometries such as rings and accretion disks stave off for as long as possible further gravitational collapse.

As mentioned above, the spiral galaxy is a particularly pertinent example of the above hypothesis. If the neutrino is the gravity bearing particle – and known to be chargeless – then it travels in essentially in a linear fashion, and Cartesian space plus time is adequate to map its path in the universe. For this reason a shape such as a spiral galaxy, with vast regions of modest curvature emanating out of a core as curved tentacles in a spiral fashion, are favorable relative to gravity because neutrinos incident perpendicular to the tangent of the curvature would only travel through relative small regions of mass occupation, and virtually no neutrinos would travel through each spiral tentacle because they are curved. So the vulnerability of survival of the spiral galaxy's "mass residents" against gravity would mostly be in the center region of the galaxy's core where a black hole would be a strongly possible consequence, and indeed near the center of our own galaxy appears such a state of highly densified condensed matter.

Because of the expectation that neutrino fluxes are not omni-directional near the expanding edges of the universe (perhaps several billion light-years from the universe's central region), it is expected that the properties of gravity are different in that region.

From our experience that the *macroscopic* is sometimes a revealing forecast of what the physical appearance of the

*microscopic* may indeed be, we might predict that the macroscopic may also be revealing "microcosm" of the "superscopic".
Such a comparison suggests that the topology of the current universe (as observed from a "distant" coordinate x,y,z,t,) may resemble that of a thermonuclear fireball at some relatively short time (perhaps 1-10 ms) after detonation. At such a time the "bang" is very successfully resisting gravitational momenta transfer from the high density flux of neutrinos into which it is expanding.
A 10 ns photograph of such a thermonuclear event, 1 ms after detonation, such that its expansion is successfully resisting gravity, shows the spatial regions that are devoid of mass and 2-d regions that would be expected from the induced-gravity model (See *Physics Today*, December 2000, p52, article by M.F. Moynihan). The time duration of this topology relative to other parameters, when normalized and proportioned, may reflect a prediction of the stamina of a typical universe before being eventually destroyed by gravity in a reverse effect, terminating with a "big thud.". The major source of the gravity bearing particles seems to be the violent "death" of previous universes of which there must have been many.

*4. Comparison to General Relativity*

Although Einstein's theory of General Relativity never claimed to explain what gravity is but instead what gravity does (beyond Newton's development), the physical explanation for gravity causing the bending of light may actually reside in the weak photon-neutrino scattering interaction. Scattering cross-sections for this interaction have been reported in the literature to be extremely small such as of the order of $10^{-70}$ cm$^{-2}$, however, the calculations were reported over 50 years ago long before the neutrino was experimentally established to have mass. The interpretation of the conclusion of the bending of light by mass should not be construed to indicate (as frequently presented in the popular press) that space is "curved", "curvilinear", or "warped", but that in any observation of mass in space through the use of light, the fact that light is bent by mass, or more precisely, deflected by the gravity-bearing particle must be considered. [A spin-off concept from this model would then be that the law of refraction's basis-- the propagation of light at different velocities in different media – must be based on on gravity-particle (perhaps neutrino) capture and scattering processes that indicate a unique graviton interaction in a particular material and in a particular phase or crystallography or morphology of that material]. That elementary particles travel faster in water than does light must be considered when constructing a fundamental theory of refraction.

In our induced gravity model, ultimate and final collapse to a black hole (associated with masses > 10 solar masses) is inferred to be due to a triggering of ultimately a coulombic (rather than gravitational) collapse (1). Our model would then explain the engulfing of light in terms of collisions between a dense flux of neutrinos with light at the ultra compressed converging regions of the forming black hole near the center of the accretion disk. The inability of light to escape is described by total absorption phenomena.

The accretion disk that is developed en route to black hole collapse is another example of predisposition toward laminar geometry, even at the extremities of nature. This indicates that the probability for a nucleon to be in the linear (or negative radial direction) path of an incoming neutrino is much higher than at normal inter-nucleon equilibrium spacing (about 1A), the latter representing the common experience of the ratio of the total volume of a crystal to mass-occupying-volume being $10^{12}$ : 1.
Under the compression of mass associated with a black hole, the mean free path of a neutrino relative to a nucleon changes from essentially the order of light years to a vastly smaller number.

*5. Relationship to the Strong Force*

For an induced gravity theory to be relevant to Einstein's and the subsequent search for a unified field theory (perhaps now better described as a unified elementary particle theory), it must fit into the new perception of the standard model. Although the research has been stopped at CERN (that had reached energies of 0.1 Tev), aimed at determining whether the Higgs-boson (theorized to engulf massless particles and create mass) actually existed, nonetheless, such an "ultimate parent particle" would not necessarily explain the roots of gravity, but instead the origin of what science refers to as mass. To establish the cause of gravity within the modified standard model, and requiring to interrogate the resulting particle system when strong meson forces are shattered by very high energy collisions, demands the acquiring of vast knowledge of neutrino collision properties under a wide range of initial conditions –a subject also ideally suited for research laboratories such as CERN and the Fermi Laboratory – and measurements of neutrino incident fluxes and densities. Indication indeed that the standard model (which does *not* account for gravity) must be modified is shown by the very recent work at Brookhaven National Laboratory where accumulated muon data is reported to be *not* in accord with predictions of the standard model. Also work that suggests that the neutrino is not affected by the strong force seems in accord with what would be expected if the neutrino is the gravity-bearing particle.

*7. The Origin of a Magnetic Field*

The identification of three different types of neutrinos that act independently with their associated quarks and leptons may point to the interpretation that the magnetic properties of each neutrino could govern the probability of the absorption or capture of the neutrino of appropriate magnetic moment by materials with missing d electrons or missing f electrons. This seems plausible because if there does not exist a gravitational field (if gravity is indeed explained by a momentum transferring particle) then one must ask why should there exist a unique "field" that is supposed to represent each of the other three long-believed basic forces? The tau neutrino has a magnetic moment *three orders of magnitude higher* the magnitude of that of its sister neutrinos. Such a neutrino may indeed pass through, generally unscathed, non magnetic materials, but interact in a particle-momentum transferring (push) manner with materials having an adequate and commensurate magnetic moment (and a positive magnetic susceptibility). [In such an hypothesis, the perfect diamagnetism associated with superconductivity's Meissner property of magnetic *repulsion* would tentatively seem to require proposing a property of a specific type of neutrino that has a unique interaction with filled orbit diamagnetic materials, causing an effect that establishes the mutual repulsion].

8. Time Dependence of Gravitational Interaction

There are a number of experiments that if performed with great care and resolution of data that could verify the induced gravity such as the expected higher value of g at midnight rather than at noon (because of the position of the sun), and an anomaly in the value of g at the time of planet syzygy that includes at least Sun, Jupiter and Mars (such a syzygy occurring May 10-May 18[th], 2001, peaking at 18 May 22:02:36 UT). However, since the Newtonian and General Relativity descriptions of gravity do not employ a time-dependence, and are field theories relating force at a distance, and since the momenta-transfer induced gravity model depends upon the capture of the gravity-carrying particle (perhaps the neutrino), and also can be calculated in terms of an impact property (delta time), then an experimental proof of time dependence of gravitational interaction would certainly favor the induced gravity model.

In the inelastic collision between a neutrino and a mass-body-nucleon, there must be time-decay or relaxation time associated with the release of the neutrino from the nucleon's quark. Since quarks are capturing (absorbing) and releasing neutrinos continuously, there will be a distribution time for this relaxation as in a collective effect (in solid-state electronics this distribution time refers to the capture of an electron by a positive hole, and is referred to as the diffusion length lifetime, of the order of 10 ns, and determines whether a semiconductor is of the lifetime type as in silicon, or the relaxation type as in gallium arsenide). For the neutrino release-from-capture this lifetime property is much longer. In the case of the mu neutrino (of about 0.2Mev to 2.0 Mev), the relaxation time is greater than a millisecond and less than a second. Therefore, if induced gravity, via momenta transfer through elastic and inelastic collisions by neutrinos, is the fundamental explanation of gravity, then a sensitive experiment should be able to detect a time dependence. Such an experiment performed in our laboratory is now described and interpreted.

Description of Experiment

Two independently suspended steel spherical pendula (of 22.35 gm and 88.26 gm) were configured 0.1 mm apart (between their adjacent circumferences), and are schematized in Fig. 1. Every experiment was performed on an optical air table to minimize extraneous vibrations. The smaller pendulum was suspended from a Newton cradle apparatus which was placed on a scale balance (triple beam as well as electronic digital balance). The center of mass (c.o.m.) of the larger pendulum is placed slightly higher (about 0.5-0.7mm) than that of the small pendulum. Each pendulum had two taut nylon suspending cables with a small acute angle separating them, hence were free to move in the x-z plane but heavily constrained or damped in the y-direction. The mass of the small sphere and its support stand from which it is suspended was measured when the large sphere was not in its vicinity (>40 cm away). The period of the small sphere was then measured under these conditions as 0.73 seconds, its very slight oscillation (through 1-3 degree of arc) being simple harmonic (SHM), and due to those continuous building vibrations that were able to be transmitted to the system via the three bolt supports of the air table.
.

The large mass was then moved to 0.1 mm from the small sphere. The balance then recorded a net loss of measured mass of about one part in 4500, this loss being due to the net attraction of the small sphere by the large sphere. The direction of this attraction is in a line a few degrees above the x-y plane because the cable of the small pendulum, not changing length, forces the line of centers of the pendula to be at a slightly elevated angle relative to the x-direction. (see diagram of Fig. 1), hence yielding a net force in the positive z direction derived from the resolution of the gravitational interaction between the pendula; this net force opposing the earth's gravitational force in the –z direction on the small pendulum, and causes a lower mass to be measured by the balance]. The small pendulum then oscillated with a period of 0.92 sec and often showed a small dwell-time or kink in the motion when closest (but not touching) to the large sphere at one extrema of its motion – its motion no longer being smoothly simple harmonic. The large sphere appeared to the unaided eye as virtually motionless throughout the period of the small sphere. The dwell time appeared, by the unaided eye, to be of the order of 0.1 s.

A neon laser was employed to be certain that there was physical space between the circumferences of the two pendula – its transmitted beam being incident on a light-detecting-diode, the output of which was directly fed into the input of an oscilloscope, and into a chart recorder on the Y- t mode. The aberration or perturbation of the motion of the small pendulum, now a forced harmonic oscillator, was further accentuated by studying the far-field scattered laser light (red) as a function of each cycle of the small pendulum, as well as by studying the intensity of the laser light incident on the light detecting diode. Occasionally, over the course of a day the motion of the two pendulum moved as a system, then released into two motions, and on some occasions the large pendulum "captured" the small one. The motion of the small pendulum lasted indefinitely when the gravitational interaction in the x-y direction exactly equalized or compensated the earth's gravitational interaction in the x-z plane.

The extremely close proximity of the two spheres was chosen because we were aware that Newton's gravitational equation inverse square law had not been verified for the range of subcentimeter-submillimeter separation distances. In our experiment, provided the inter circumferential distance between the two pendula at closest proximity is of the order of 0.1 mm or less, and the values of g are calculated for the SHM period of the small sphere, and for the perturbed period of the small sphere (perturbed by the large sphere), and the gravitational force of attraction between the two pendula spheres determined, it is shown that the calculation of r from the inverse square relationship gives a value much less than the inter-center distances of the small and the large pendula, in fact a distance (of the order of 0.01- 0.04 mm) even less than the order of the separation distance of the circumferences of the two spheres (0.1mm), therefore challenging the concept of a central force notion associated with the hypothesis that all mass can be equivalently thought to exist at the center of mass in this interaction (valid only from integration using an inverse square relationship). The data is, however, well suited to $r^{-4}$ dependence (r calculated to be 0.35 mm) if the Cavendish value of G is employed (however, it is not apparent that G would still be of the order of $10^{-11}$ in the submilllimeter regime, yet G is utilized in expressions for gravitational potential which include correction factors for very small r; see Ref. 4). The higher power representing n=2 in the theoretically predicted exponent expression 2+n. would yield for an $r^{-4}$ dependence (via integrating) pertinent shells of mass near the circumference of the pendulum, in accord with what would be more reasonable in a particle capture momentum-transferring model for gravity. The $r^{-4}$ dependence could be interpreted by string theorists to suggest that the n=2 represents two large extra dimensions.

We were subsequently made aware (Ref 3) that recent work has been reported (4) to be interpretable through Newtonian gravity at separation distances down to 197 microns. In Ref 4 are given references that suggest that the unexplored short-range regime of gravitation (length scales below a few mm) may yield a gravitational interaction that could display fundamentally new behavior (5-8). In Ref 4 is also given a "phase field" diagram which plots parameters that relate to variation from Newton's gravitational equation. The diagram gives a region that relates to approximately separations between the order of 0.1 mm and 0.15 mm, referring to this region as the "vacuum energy limits" in which the variation from Newtonian gravity is sharply changing as separation is further reduced. It is also noted in Ref 4 that the gravitational cosmological constant ($3 keV/cm^3$) that is deduced from distant Type 1A supernovae (Refs. 9, 10) corresponds to length scales of 0.1 mm. This and other work suggests that the Newtonian gravitational potential be modified to the expression:

$$V(r) = - G[(m_1)(m_2)/r] (1+ alpha\ e^{-r/lambda})$$

Where the aforementioned "phase field" diagram is a confidence upper limit plot of the absolute value of alpha versus lambda relating to inverse square law violating interactions. From our own work (1) the value of G, above, would depend upon the scattering cross-section of the gravity-bearing particle with the constituents of nucleons, therefore also with any parameter that affects that cross-section such as temperature, pressure, and geometry.

The reason for the above indication of r being less than the inter-center distance is two-fold. Theoretically, it is doubtful that the exponent of r could be possibly 2 at such short proximities, and certainly not at the distance between the proton and the electron cloud of an atom. [Newton derived the inverse square postulate by using Kepler's First Law of (elliptical) Planetary Action in which r is extremely large, and tested his theory employing the moon-earth distance. Cavendish determined G from a laboratory experiment of small r, but much larger than 0.1 mm]. Thus the exponent of r may possibly be changed from 2 in the Newtonian gravitational force postulate to probably 2+n in the close-proximity system. However, more important, if an external particle theory of gravity is fundamental, then the cross-section of each pendulum that captures the greatest number of neutrinos from the unshadowed direction (away from the other pendulum) per unit time is not physically likely to be the polar cross-section (perpendicular to the equatorial plane) but a time-averaged cross-section closer to the circumference of the sphere – this due to the high kinetic energy and long mean-free-path of the low-mass neutrino. This then seems to be a reasonable interpretation of the data.

The 88.26 gram stainless steel pendulum was then replaced with a 1.30 kg singly-suspended iron pendulum of much larger surface area. The perturbed period of the small pendulum then increased to 1.38 sec, and a frequent dwelling of the motion was again detected. Table 1 gives the pertinent data for the above experiments which were repeated over 200 times.

Table 1

| Pend | $T_m$ (s) | m | Mij | (m)Mi,j | S.A.(M) | g(eff) | mg–mg(eff) | d$T$ |
|---|---|---|---|---|---|---|---|---|
| | Sec | gram | gram | gram$^2$ | cm$^2$ | cm sec$^{-2}$ | N | sec |
| m | 0.73 | 22.4 | 0 | 0 | 12.6 | 800 | | 0 |
| mMi | 0.92 | 22.4 | 88.0 | 1971 | 24.6 | 503 | 6980 | 0.19 |
| mMj | 1.38 | 22.4 | 1840 | 41,216 | 186.2 | 224 | 13536 | 0.65 |

g(eff) is calculated by $T = 2\pi (L/g)^{0.5}$

mg-mg(eff) based on g that is calculated from the experimentally-determined period of m for deflection of the pendulum of less than 2 degrees.

d$T$ = *change in period, T*

The graphs of d$T$ vs mMi,j, d$T$ vs surface area of Mi,j, and $T$ vs mg(eff) are given at the end of the paper in Figs 2a,b and 3. In Fig 2, the graphs show the character of the rising portion of a parabolic curve, indicating that the period of the lower-mass pendulum will rise as the mass of the second higher-mass pendulum increases, but at a decreasing slope. In Fig. 3, the dependence appears hyperbolic, indicating that at very high masses of the higher-mass pendulum the period of the lower-mass pendulum will increase until its motion stops and it resides in contact with the higher mass pendulum such as a body in contact with the earth, as expected classically.

**Fig 4 shows that the dwelling effect in the form of the occasional *flattening* of only the part of the Y- time chart-recording (taken at 2 sec/cm) of the output of the light detecting diode corresponding to the minimum intensity level of the laser when the two pendula were at closest proximity (the up-direction in the chart recording). This flattening occurred approximately once out of every ten cycles of the small pendulum, and the flattened region was approximately of one-eighth of a second duration.**

Mathematical Analysis of Time-Dependence and Dwell Time

Referring to the diagram in Fig 1, the following assumptions are made: [where **p**(nu1) refers to the momentum of neutrino # 1, **p**(nu2) gives the momentum of neutrino #2, m = mass of steel sphere struck directly from the atmosphere by nu1, M = mass of the steel sphere struck directly from the atmosphere by nu2, t1(rel) = release time of neutrino #1 by nucleon (quark); t2(rel) = release time of neutrino #2 by a different nucleon, u refers to u-quark, d refers to d-quark, p refers to proton, n refers to neutron, nu $_e$ refers to electron-neutrino, and <u>nu</u> $_e$ refers to electron-antineutrino, e $^-$ refers to electron, e$^+$ refers to positron) i). **p**(nu1) is taken out of the neutrino flux 1, for time t = t1(rel) because of capture by m, and would otherwise be incident upon M, but instead pushes m toward M for some time interval.

**p**(nu2) is taken out of flux 2, for t=t2(rel), because of capture by M, and would otherwise be incident upon m, but instead pushes M

toward m for some time interval.

The capture (inelastic collision) given by: $p(u,u,d) + nu_e \rightarrow n(u,d,d)$
Release given by: $n \rightarrow p + e^- + \underline{nu}_e$
and by $p \rightarrow n + e^+ + nu_e$

ii) for nu1 and nu2, the component velocities, $v_y = v_z = 0$.

iii) Assume nu2 (in ii, above) is transiently-captured at coordinate $(X_M, 0, 0)$ in M, and nu1 (in I, above) is transiently-captured at coordinate $(-X_m, 0, 0)$ in m, and assume for the moment no further capture of nu2 by M, or nu1 by m.

Let td2 = time until nu2 momentum is eventually transferred to m, relative to some $t_o$, similarly for td1, and assume that nu2 has then penetrated m up to $(-X_{cm}, 0, 0)$ before transferring its momentum inelastically to m; similarly for nu1 and M at $(X_{cM}, 0, 0)$.
It is assumed that t=0 when a nu2 arrives at circumference of M, and a nu1 simultaneously (or almost so) arrives at circumference of m; dM = diameter of M; dm = diameter of m;
R=distance from x=0 to point on circumference of M nearest x=0;
r=distance from x=0 to point on circumference of m closest to x=0.

It is assumed that the velocities of nu1 and nu2 are the same as that of light, c. Let $t_{2m}$ = response time for nu2 momentum to be imparted to m, after nu2 penetrates m until capture point, this due to consideration of any degree of "softness" of nucleon, and time dependence of final capture converting a u-quark to a d-quark); similar for $t_{1M}$.

Then,
td2 = $\{[(d_M + R) - X_M]/c\} + t_{2r} + [X_M/c] + [X_{cm}/c] + t_{2m}$

During $t_{d2}$ there is a shadowing of impinging momentum from the + x direction that would otherwise displace m toward the – x direction. If within time $t=t_{d2}$ a gravity-bearing particle (appropriate neutrino) from the direction of nu1 is captured by m, and transfers its momentum inelastically to m, then m is displaced toward M during that time because m is **not** counterbalanced by inelastically absorbed momentum from nu2.

Therefore, gravitational interaction is not based on a field, or a force-at-a-distance interaction, but on the statistical imbalanced transfer of particle momentum. This statistical property is shown in Fig 4 reflecting that a cumulative delay event appears to take place irregularly (probably essentially randomly without unique harmonic frequencies) on the average of once out of every ten cycles of the pendulum.

In relation to the dwell-time of the pendula in the experiment, the "dwell" is due to the net or collective effects of $t_{d2}$ and $t_{d1}$ When t exceeds some critical value for a particular statistical distribution of the independent variables, then the transient pushing-together ("attraction") of the two pendula (approximately represented by m and M, above) is overcome by the gravitational particle interaction in the z direction (gravity between m and earth, and between M and earth).

The significance of the 0.1 mm distance may be indication of the wavelength of the pilot wave that is associated with the gravity bearing particle (graviton). A separation distance that is less than the wavelength of the graviton (possibly neutrino) would give rise to anomalous or unusual scattering effects. These are analogous to those predicted for superlattices in epitaxial layering structures whose repeat distance is larger than the wavelength of the light which is incident upon the structure (originally predicted to lead to

Tz oscillation). The 0.1 mm wavelength suggests that the mass of even a cluster of in-phase gravitons is extremely small, and that the frequency of the pilot wave is about 3 Tz.

9. Experiments on Temperature Dependence of Weight

It is reasoned that if gravity is due to momentum transfer from an elementary particle, then any phenomena that increases the probability for the capture of such a particle should cause an increase in measured weight. Such a phenomena is an increase in heat which will cause an increase in internal energy in the form of increased amplitude of atomic or molecular vibration that should give rise to a greater collision cross-section.

This hypothesis was tested in the laboratory by utilizing a laboratory weight designated as 500 grams, and accurately measuring its mass at 25C and then at 100C, and then measuring the mass as a function of cooling-time. The brass test weight was heated in boiling water then extracted remotely, dried quickly and placed on the laboratory balance in exactly the same location where it was weighed before immersion into the water. No further tampering was performed in on the balance in Trials 1 and 2 in Table 2 (given at the end of the paper), except the dial movement to create equilibrium. These data are given for three of 20 equivalent-results separate trials in the Table, and when plotted indicate a temperature coefficient of measured weight of +1.3 dynes per degree centigrade. In the third trial in Table 2, the balance was re-zeroed between every measurement at precisely 15 minute intervals. These results suggest that what has in the past been believed to be a random error of measurement of mass (thus, weight), even at constant temperature, may be at least in part due to an anomaly or a fluctuation in the density of the gravity-bearing particle flux of approximately one to 10 parts in $10^3$.

10. Velocity of the Gravity-Bearing Particle

The detecting of the neutrino by super K in Japan, and the inference from the statistics of up-going and down-going electron-neutrinos and muon-neutrinos of it having mass employs the consideration that a the neutrino travels faster in water than does light in water [index of refraction (n) for white light in water =1.33; v (light in water)=$2.25 \times 10^8$ m/sec] This consideration is consistent with the requirement that the gravity bearing particle must be relativistic, and in a material medium other than the neutrino flux itself, the gravity bearing particle's velocity, between capture collisions, is unimpeded because it is not acted upon by gravity unless it is scattered in rare instances by another gravity-bearing particle. Since for air of the earth's average atmosphere, n=1.0003, then one could argue that the neutrino penetrates the earth's atmosphere at a velocity much greater than $2.25 \times 10^8$ m/sec, and possibly as high or higher than light's velocity in air ($2.997 \times 10^8$ m/sec). In the candidacy of the neutrino as the graviton, the collision cross-section for a neutrino-neutrino collision has been calculated to be of the order of $10^{-70}$ (see references within Ref. 1). On the other hand, the photon is more effectively impeded by gravity and its propagation attenuated due to scattering. The source of the great majority of impinging neutrinos is so distant, in excess of perhaps $10^{50}$ meters that despite its mean free path being calculated as of the order of light years (1 light year = $10^{15}$ meters), by the time a flux bundle reaches earth, there are abundant opportunities for inelastic collision when nucleons are encountered in its path.

11. The Cosmological Constant

The Cosmological constant is a repulsive term introduced by Einstein in 1917 in order to compensate for the effect of gravity such that the net result will give a static (unchanging with time), closed, unbounded, finite (neither expanding or contracting) universe, yet eventually discredited by Einstein, and by others when the Hubble work indicated through the red shift an expanding universe. A Cosmological constant of zero is then inferred to indicate that the vacuum energy density is zero (implying the proviso that gravity is negligible), and conversely a non-zero Cosmological constant implies a nonzero value of the vacuum energy density. Gravity, however, is not negligible in cosmological physics, and the inability of physics heretofore to explain the extremely small magnitude of the vacuum energy density is considered by many particle physicists as one of the major problems in modern physics (11).

The correlation in our work of our measured critical separation length of 0.1 cm with the vacuum energy density of
$\Lambda = 3$ keV/cm$^3$ {corresponding to length scale $= [hc/(\Lambda)(2\pi)]$ suggests that the Cosmological constant is not zero but comparatively small. Our external particle model for gravity readily gives rise to a repulsive interaction between mass bodies (nucleons) that are extremely far apart (such as near opposite extremities of the universe), because those gravity bearing particles whose sources are *within* the existing universe would have a "shadowing" effect opposite to what is described within the bulk of the universe in Ref. 1; this would offset balanced effects of gravity particles emanating from outside the universe. Sources of gravity-bearing neutrinos from within the universe are supernovae explosions and stars such as the Sun, and at a critical huge length scale, most of these astronomical bodies would be located *between* the two said nucleons rather than exterior to their lines of centers. Since the source of gravity bearing particles within the universe would eventually expire, it would appear that the Cosmological constant is becoming smaller, and that depending upon sources of neutrinos, the end of this universe could be in the form of a gravitationally-induced collapse. (From inferences regarding the density and properties of supernovae and stars, this time period to a possible collapse should be able to be calculated to orders of magnitude).  The above-cited energy density refers to the region of alpha vs lambda in Ref. 4 Fig 4 bordered by the length scale of  0.1 mm and referred therein to as vacuum energy limits.  Our results now indicate then another real surfacing of an indication of  a current non-zero Cosmological constant.
Based on the above, it appears that this experiment could arguably be interpreted as an indication of the detecting of the graviton.

Acknowledgments

Our study of the fundamental cause of gravity was an outgrowth of discussions with colleague of 37 years, theoretical physicist, Dr. William Stanley, regarding my own contention that no physical phenomenon could explain the existence of a Newtonian central gravitational force, and hence questioning the existence of a Newtonian gravitational field, and our interest to explore the gravitational behavior in the enormous void space between nucleons in a solid.  Dr. Stanley's postulating of the neutrino as the omni-directional particle subject to "shadowing" by inelastic collision with a mass body's (earth's, for example) nucleons then triggered three years of theoretical study that led to the experiments described above.

The author wishes also to acknowledge the able laboratory assistance and keen observations of:  Eileen Leung, Max Felsher, Nevi Kosa, Jennie Blodgett, Chris Kocur, Matt Pooley, Andrew Magliozzi, Dershen Patel, Olivia Zurek, Danielle Bouldoc, Gina Tassone, Dave Hurley, Jake George, Ray Chen, James Battaglia, and Dave Malfroy; the careful bibliographical work conducted by Kitty Marie Hurst Vezzoli; discussions with Dr. F. Zerilli thirty years earlier regarding his mathematical physics work on gravity waves when at Princeton University with Professor John Wheeler; and the loan of experimental equipment by Professor Larry Rubin, Emeritus, at The Massachusetts Institute of Technology, where the author of the current work was a Visiting Scientist and a user of the National Magnet Laboratory facility from 1987 though 1994; and Brian Van Konynenburg for internet support.

Table 2

| Time | Temperature (C) | Mass (grams) | |
|---|---|---|---|
| 1030 hrs | 22.0 | 499.83 | |
| 1100 | 99.5 | 499.93 | |
| 1112 | * unm | 499.86 | |
| 1220 | * unm | 499.82 | |
| 1342 | *22.2 | 499.81 | |

Trial 1 (above) and Trial 2, 3 (below) Data:
Heating in water, and dry cooling (*) in air.
unm: unmeasurable without causing intrusion

| Time | Temperature (C) | Mass (grams) | |
|---|---|---|---|
| 1338 | 22.8 | 499.84 | |
| 1358 | 99.5 | 499.90 | |
| 1405 | * unm | 499.90 | |
| 1435 | *unm | 499.89 | |
| 1442 | *unm | 499.87 | |
| 1505 | *unm | 499.87 | |
| 1630 | 23.0 | 499.86 | |
| | 18 | 498.81 | |
| 1000 | 100 | 498.95 | |
| 1020 | *unm | 498.90 | R.T.= 20.0 C |
| 1040 | *unm | 498.92 | R.T.= 20.0 C |
| 1109 | *unm | 498.89 | R.T. =19.1 C |

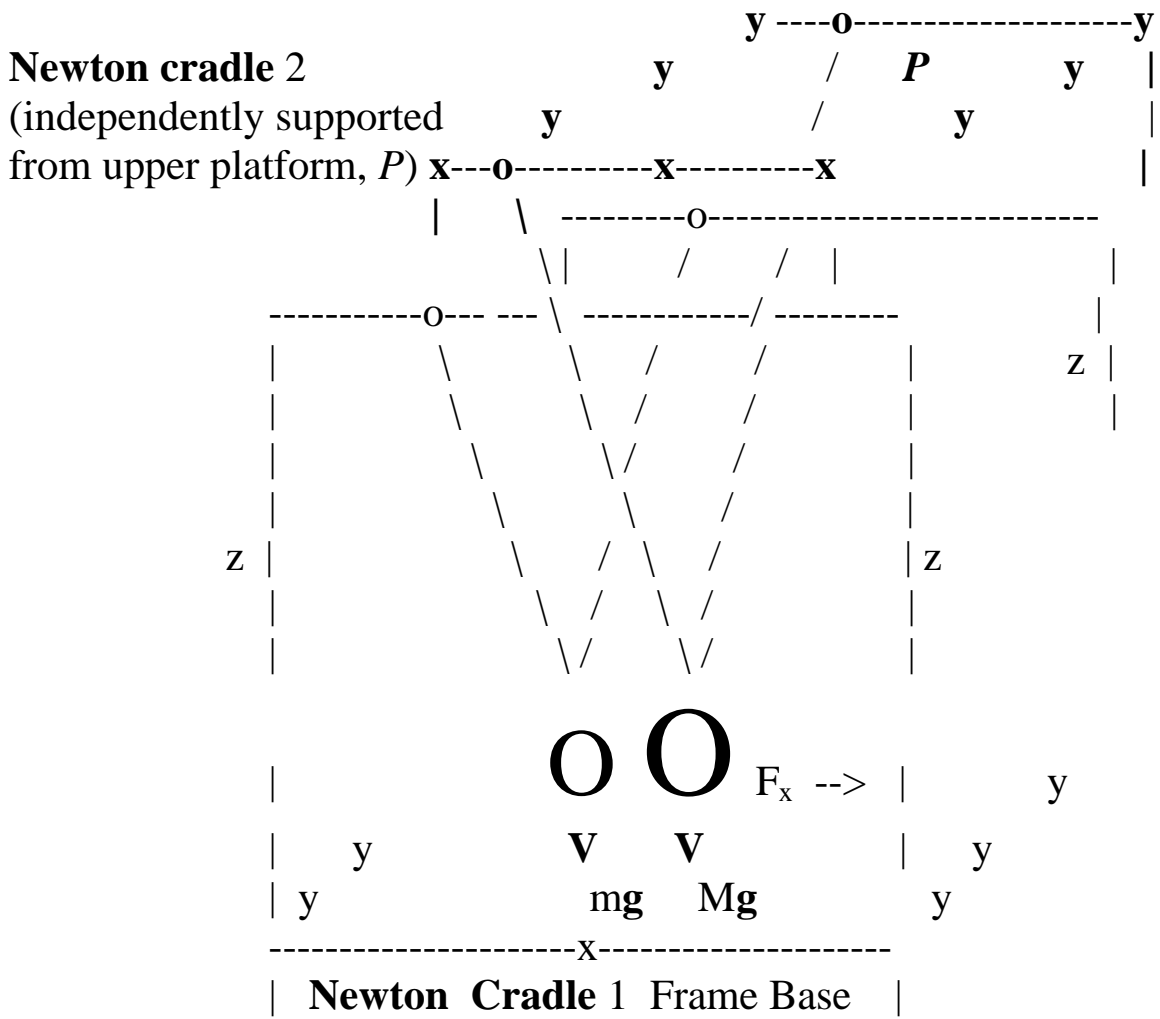

Fig. 1 Schematic for Two Newton Cradle Pendula
Experiment to Study Change in Period of
Motion and Dwell Due to Gravitational
Interaction between m and M.
(Oscillation in the x-z plane above, heavily damped in y).
$t_{m-M}$ indicates the period of m when M is removed;
$t_{m+M}$ indicates period of m (longer) when M is very close,
R + r = distance between m and M; r+R << 1 mm
(o and **o** : suspension points of cables for m and M)

Figure 2

Graphs of Delta Tau vs, Products of Masses of m and Mij,
and vs Surface Area of Mij

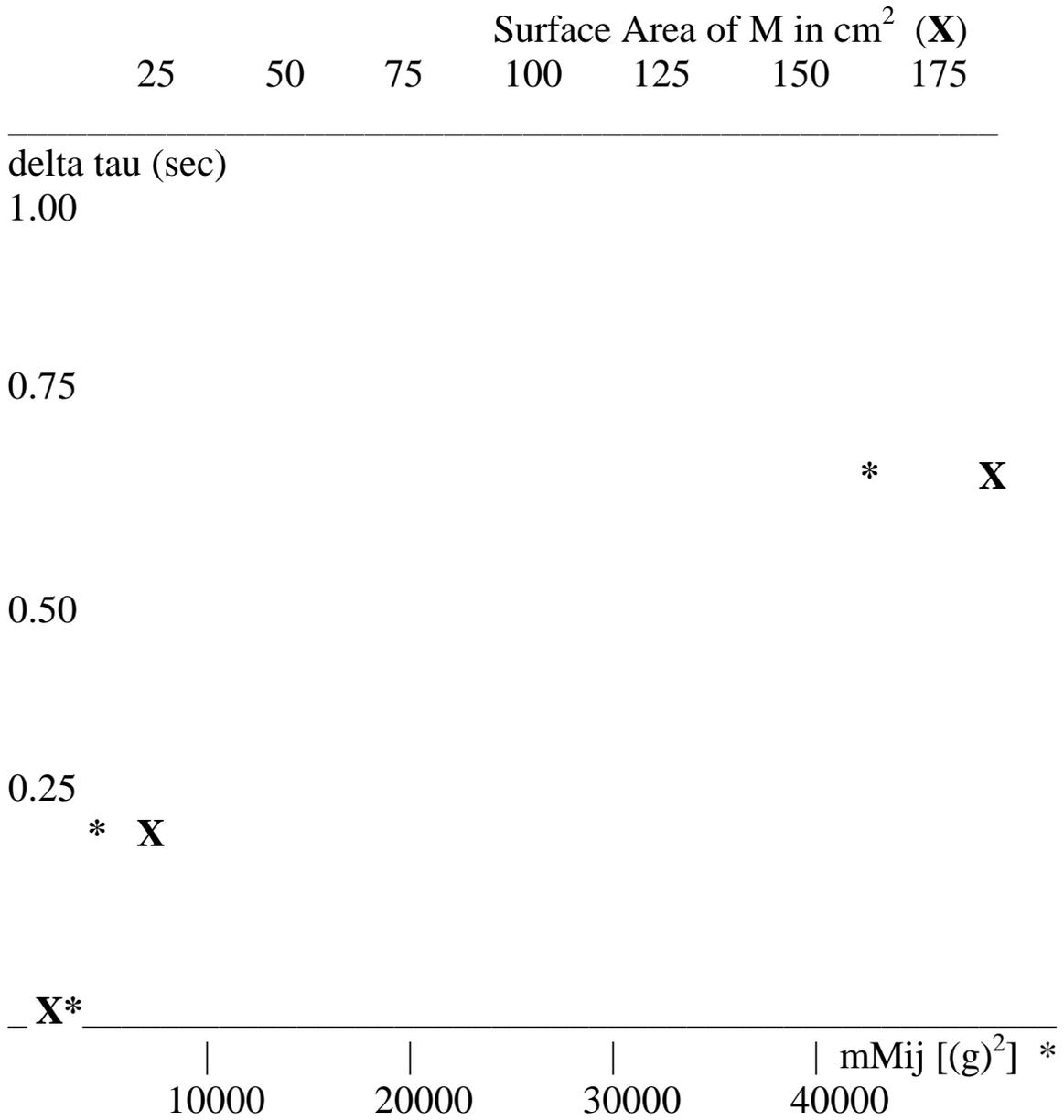

Surface Area of M in cm$^2$ (**X**)
    25      50      75     100    125    150    175
________________________________________________

delta tau (sec)
1.00

0.75

                                      *    **X**

0.50

0.25
  * **X**

_**X***________________________________________
      |       |       |    | mMij [(g)$^2$] *
  10000  20000  30000  40000

(at the origin **X** and * are superimposed)

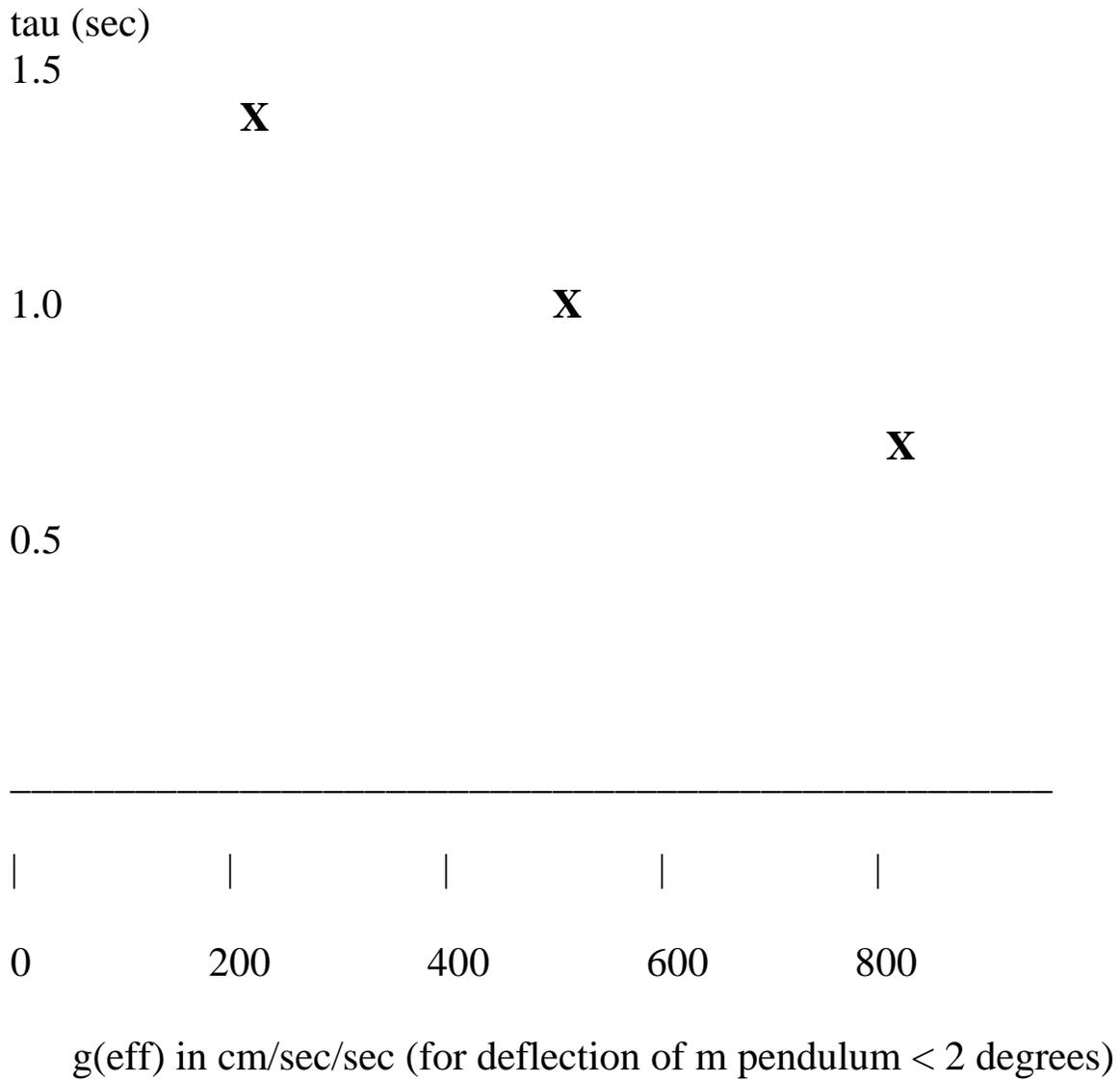

Fig. 3
Graph of Period (tau) in sec vs g(eff) cm/sec/sec

g(eff) in cm/sec/sec (for deflection of m pendulum < 2 degrees)

# Submillimeter Length Scale Time Statistics of Gravitational Interaction

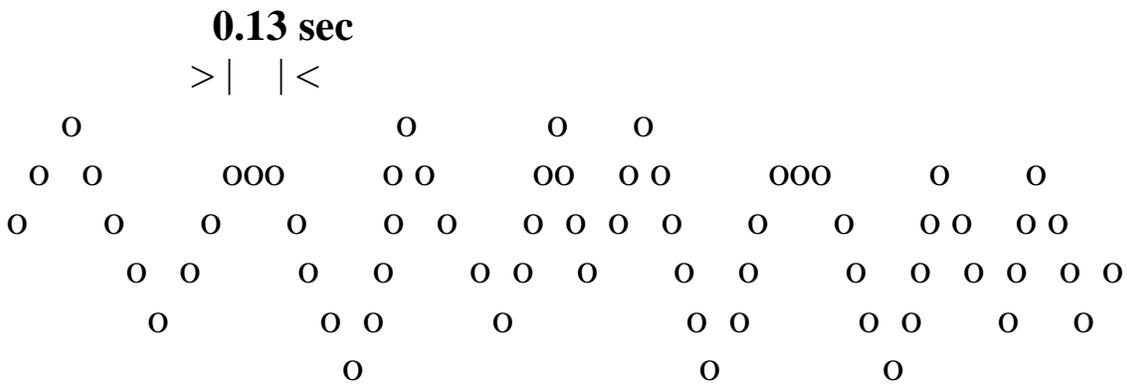

Low output of laser-light detecting diode direction.
(corresponds to steel pendula at minimum separation)

Flattened or truncated crests indicate dwell time of small mass pendulum in close (<0.1mm) to large mass pendulum

Chart speed = 2 sec per centime
(data digitized from analogue X-Y recorder raw data)
High output of laser light detecting diode direction.
[corresponds to steel pendula at maximum (2mm) separation].

36 flattened crests out of first 300 cycles of the above experiment.

Fig. 4. Output of laser-light detecting diode (input of Y-t chart recorder) as a function of time at 2 sec/cm, showing flattened crests of wave train associated with pendula motion, and indicating dwell-time of small mass pendulum in close proximity (<0.1mm) to large mass pendulum. Experiment indicates time-dependence. of perturbation of small mass by large mass, and consequently a time property of the source of gravity.

References:

1) W. Stanley and G. C. Vezzoli, Induced Gravity Model Based on External Impinging Neutrinos: Calculation of G in terms of Collision Phenomena and Inferences to Inertial Mass and Atomic Quantization," in Los Alamos National Laboratories Preprint Gallery, 7 February 2001. Web site http://xxx.lanl.gov//.
(In refereeing response process J. Quant. Grav.)

2) N Arkani-Hamed, Savas Dimopoulos, and G. Dvali, Sci.Amer. Aug. 2000, 62

3) Elizabeth Simmons, Boston University Physics Department, Neutrino Workshop, March 24, 2001

4) C.S. Hoyle, U. Schmidt, B.R. Heckel, E.G. Adelberger, J.H. Gundlach, D.J. Kapner, and H.E. Swanson, Phys. Rev. Lett. 86(8), 19 Feb 2001

5) R. Sundrum, J. High Energy Phys. 9907, 001 (1999).

6) N. Arkani-Hamed, S. Dimopoulos, G. Dvali, and N. Kaloper, Phys. Rev. Lett. 84, 586 (2000).

7) S. Dimopoulos andG. Guidice, Phys. Lett. B 379, 105 (1996).

8) N. Arkani-Hamed, S. Dimopoulos, and G. Dvalli, Phys. Lett. B 429, 263 (1998).

9) A. G. Riess et al., Astron. J. 116, 1009 (1998).

10). S. Perlmutter et al., Astrophys. J. 517, 565 (1999).

11). **The New Physics**, edited by P. Davies, Cambridge, Cambr. Univ. Press, 1998. See also P. Meszaros, Science, 291, 5 Jan 2001.